\pdfoutput=1
 \documentclass[aps, prd, twocolumn, nofootinbib,superscriptaddress]{revtex4-2}

\newcommand{\aap}{Astronomy and Astrophysics}

\newcommand{\jcap}{Journal of Cosmology and Astroparticle Physics}

\newcommand{\araa}{Annual Review of Astronomy and Astrophysics}

\usepackage{natbib}
\usepackage{amsmath}
\usepackage{float}
\usepackage{hyperref}
\hypersetup{colorlinks=true, allcolors=blue}
\usepackage{graphicx}
\usepackage[english]{babel}
\usepackage{xcolor}

\newcommand{\be}{\begin{equation}}
\newcommand{\ee}{\end{equation}}
\newcommand{\ba}{\begin{eqnarray}}
\newcommand{\ea}{\end{eqnarray}}
\newcommand{\LCDM}{$\Lambda$CDM}
\newcommand{\planck}{\textit{Planck}}
\newcommand{\wmap}{\textit{WMAP}}

\begin{document}

\title{A foreground-marginalized `BK-lite' likelihood for the tensor-to-scalar ratio}

\author{Heather Prince}
\affiliation{Department of Physics and Astronomy, Rutgers, the State University of New Jersey, 136 Frelinghuysen Road, Piscataway, NJ
08854, USA}
\affiliation{Peyton Hall, Princeton University, Princeton, NJ 08544, USA}
\author{Erminia Calabrese}
\affiliation{School of Physics and Astronomy, Cardiff University, The Parade, Cardiff, CF24 3AA, UK}
\author{Jo Dunkley}
\affiliation{Joseph Henry Laboratories of Physics, Jadwin Hall, Princeton University, Princeton, NJ 08544, USA}
\affiliation{Peyton Hall, Princeton University, Princeton, NJ 08544, USA}

\begin{abstract}
The current limit on the tensor-to-scalar ratio from the BICEP/Keck Collaboration (with $r<0.036$ at 95\% confidence) puts pressure on early universe models, with less than 10\% of the error on $r$ attributed to uncertainty in Galactic foregrounds. We use the BICEP/Keck BK18 public multi-frequency likelihood to test some further assumptions made in the foreground modeling, finding little impact on the estimate for $r$. We then estimate foreground-marginalized cosmic microwave background (CMB) $B$-mode polarization bandpowers. We fit them with a multivariate offset-lognormal distribution and construct a marginalized `BK-lite' likelihood for the CMB $B$-mode spectrum with no nuisance parameters, serving as a method demonstration for future analyses of small sky regions, for example from the South Pole Observatory or CMB-S4. 
\end{abstract}
\maketitle

% - - - - - - - - - - - - - - - - - - - - - - - - - - - - - - - - - - - - - -
\section{Introduction}
% - - - - - - - - - - - - - - - - - - - - - - - - - - - - - - - - - - - - - -

A key goal towards understanding the physics of the early universe is to constrain or detect primordial tensor perturbations. These would have propagated as gravitational waves, imprinting signals in the cosmic microwave background (CMB) intensity and polarization anisotropy. They uniquely produce primordial divergence-free $B$-mode polarization, with power predicted to peak at degree scales and larger \citep{seljak/zaldarriaga:1997,kamionkowski/kovetz:2016}. $B$-mode polarization is also generated from the gravitational lensing of the primordial curl-free $E$-mode polarization, and is additionally produced by thermal dust and synchrotron emission in the Galaxy \citep[e.g.,][]{planck2018_dust}.

The tightest constraint on primordial tensor perturbations comes from BICEP3 observations from the South Pole, together with data from BICEP2 and the Keck Array, and supplemented by \textit{Planck} and \textit{WMAP} satellite data, resulting in an upper limit on the ratio of tensor to scalar power at 0.05 Mpc$^{-1}$ of $r<0.036$ at 95\% confidence \citep[][hereafter BK18]{Bicep2021}. Independent of BK18, an upper limit of $r<0.056$ was estimated from \planck, using the NPIPE maps \citep{tristram/etal:2021}, and the first flight of the SPIDER balloon experiment gave $r<0.11$ \citep{spider2021}.

These constraints on $r$ depend on the treatment of Galactic foregrounds, especially the polarized dust emission. To constrain the foregrounds in the BK18 analysis, polarization maps at seven effective frequencies are fit to simultaneously constrain $r$ and a seven-parameter foreground model. In the original BICEP2 150~GHz data \citep{bicep2_I,BKP2015}, the foreground level at degree scales was significantly larger than the gravitationally lensed CMB signal expected for a universe with $r=0$. In contrast, the best-measured $B$-mode signal in BK18 is now at 95~GHz where the power in foreground emission is more than five times lower, of comparable size to the lensed CMB at degree scales. The foregrounds still need to be modeled, however, and a set of different models were tested in the BK18 analysis, demonstrating the stability of the estimated $r$ to the choices made. In this paper we briefly explore some further assumptions, finding little impact on $r$ using the publicly available products. We then use the BICEP/Keck, \textit{WMAP} and \planck\ data to estimate foreground-marginalized bandpowers for the $B$-mode angular power spectrum, an extension to the BK18 analysis. We use them to construct a `BK-lite' likelihood which reproduces the same tensor-to-scalar ratio constraint as the corresponding multi-frequency likelihood, following a similar approach adopted for data from the Atacama Cosmology Telescope (ACT) and \planck, and for simulated Simons Observatory data \citep{Dunkley2013,Calabrese2013,planck2015_likelihoods, wolz2023}. This likelihood can be used to test models of the early universe and to obtain distributions of the $B$-mode bandpowers that include uncertainty due to foregrounds.

This paper is laid out as follows. We review the BICEP/Keck data and likelihood in \S\ref{sec:data}. In \S\ref{sec:tests} we explore the foreground model. In \S\ref{sec:compress} we estimate foreground-marginalized CMB $B$-mode bandpower amplitudes and construct a `BK-lite' likelihood. We conclude in \S\ref{sec:discussion}.

\section{The BICEP/Keck data and likelihood}
\label{sec:data}

In this section we review the details of the BK18 likelihood analysis, described fully in e.g., Refs.~\cite{Barkats2014,Bicep2_2014,Bicep2021}. 

\subsection{Data}

\textbf{BICEP3:} $Q$ and $U$ Stokes vector maps were made from data gathered from 2016--18 with BICEP3, over $\sim 600$~deg$^2$ in a band centered at 95~GHz. They have the lowest polarization noise yet reported, with a depth of 2.8 $\mu$K-arcmin \citep{Bicep2021,Bicep3_2022}. 

\textbf{BICEP2/Keck:} Stokes maps were made in a smaller $\sim400$~deg$^2$ region using data gathered with BICEP2 from 2010-12 in a band centered at 150 GHz \citep{Bicep2_2014}, and with the Keck Array from 2012-2018 in bands at 95, 150 and 220 GHz \citep{Keck2015,BK15}. 

\textbf{\textit{WMAP} and {\it {\textbf {Planck}}}:} To constrain synchrotron radiation and dust emission, the 23 and 33 GHz maps from the \textit{WMAP} satellite are used, together with the 30, 44, 143, 217 and 353 GHz maps from the \textit{Planck} satellite. They are all masked using the same apodization mask as used for the BICEP3 data. \\

\begin{figure}[!t]
    \vspace{-0.2in}
  \centering
\includegraphics[width=1.08\linewidth]
  {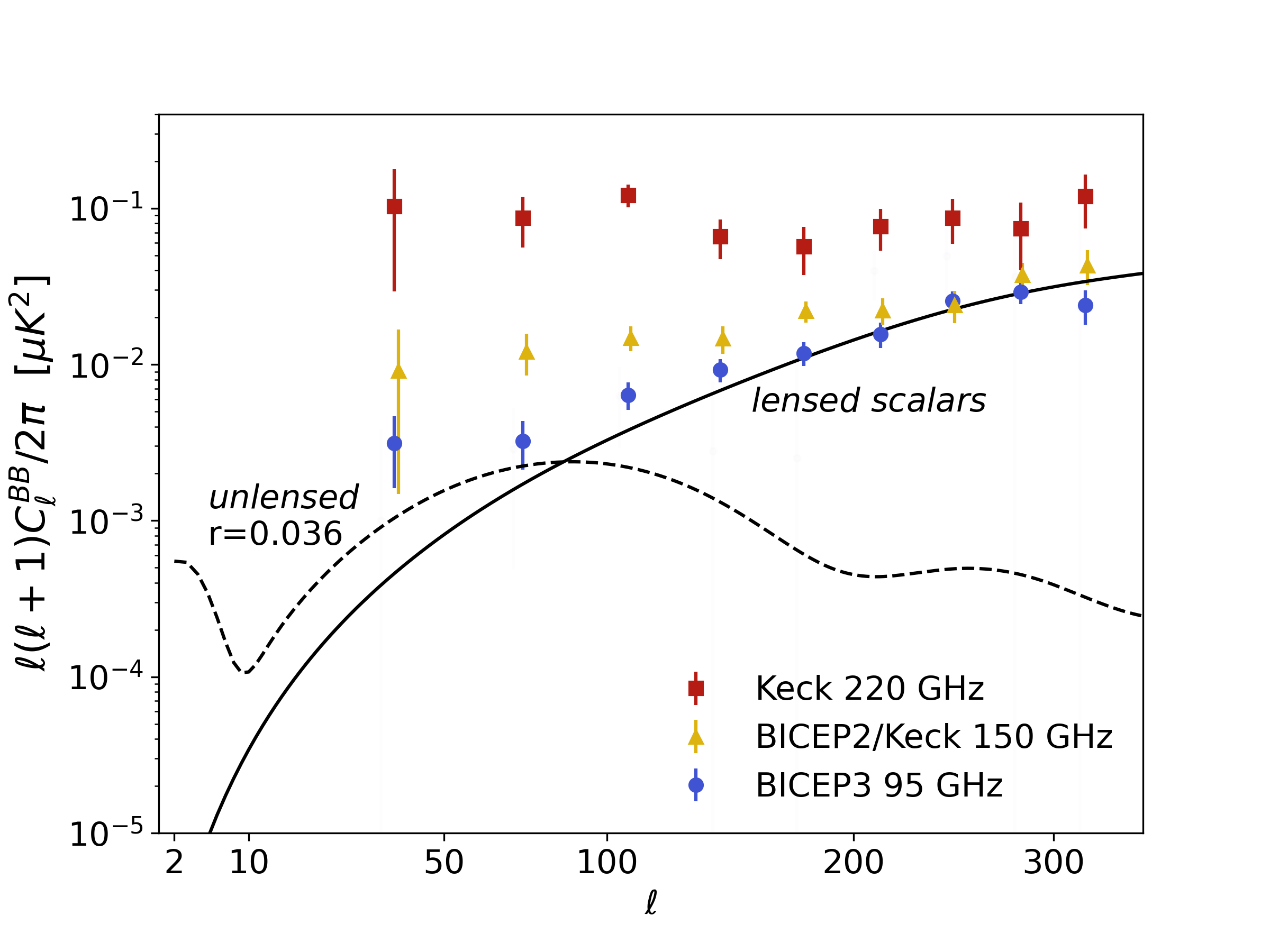}
  \caption{Power spectra for the 95~GHz BICEP3, 150 GHz BICEP2/Keck and 220 GHz Keck array data, from the BK18 public products. Also shown is the lensed scalar power spectrum for a $\Lambda$CDM cosmology with $r=0$ and the unlensed prediction for $r=0.036$ (the current 95\% upper limit). No synchrotron is detected in these spectra; the excess power over the lensed scalar signal is attributed to dust emission.} 
     \label{fig:bk_spectra}
\end{figure}

In total there are 11 polarization maps included in the analysis, effectively spanning 7 frequency bands.
This results in 11 auto-spectra and 55 cross-spectra for $BB$.  Each spectrum has 9 bandpowers in the range $20 < \ell < 330$, with $9\times 66=594$ total bandpowers, fit simultaneously in the analysis. Figure \ref{fig:bk_spectra} shows just the BICEP3 95~GHz and BICEP2/Keck 150 and 220 GHz auto-spectra bandpowers from BK18, highlighting the relatively low foreground level at 95~GHz compared to 150~GHz. The 220 GHz data are dominated by dust at all the scales shown, and the 150 GHz data are dominated by dust on degree scales $\ell \lesssim 200$. 

\begin{figure*}[!bth]
\hspace{-0.25in}
\includegraphics[width=0.9\linewidth]{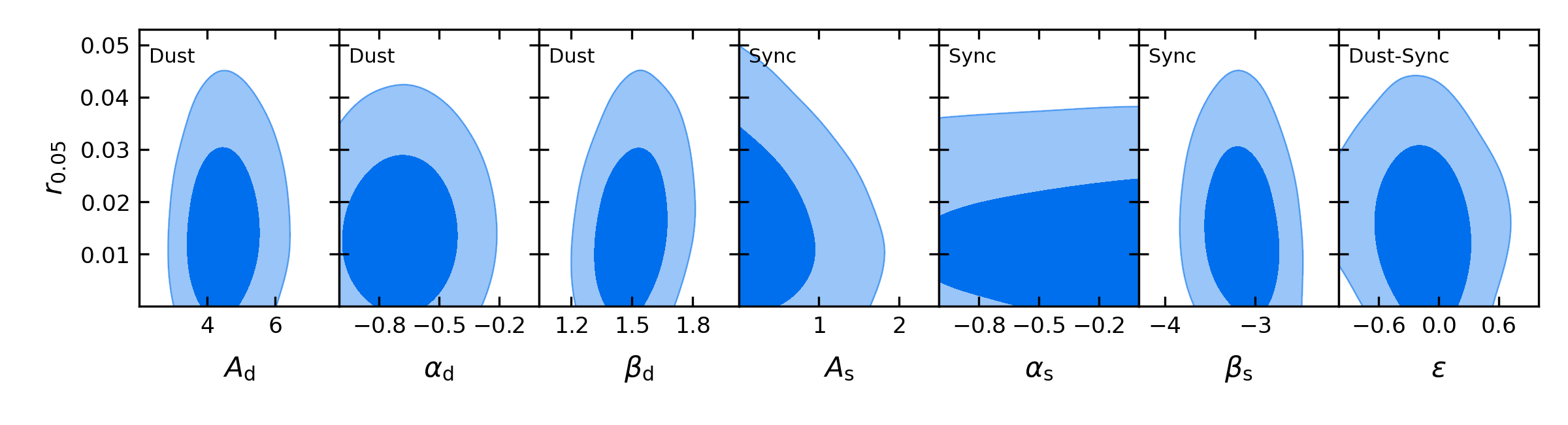}
\hspace{-0.35in}
\vspace{-0.2in}
\includegraphics[width=0.16\linewidth]{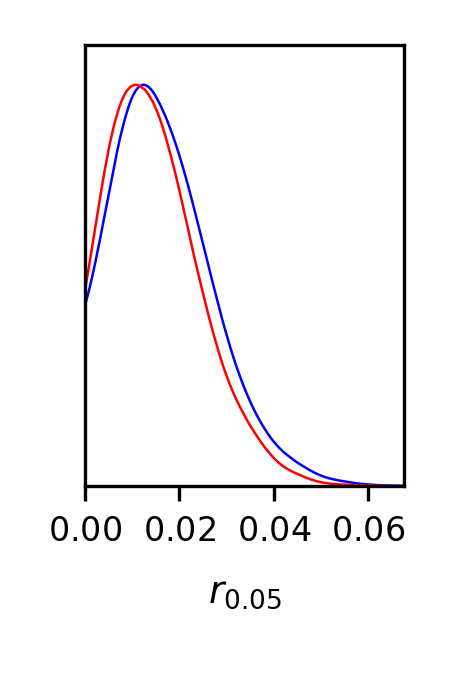}
  \centering
  \caption{Reproduction of estimates for $r$ and seven foreground parameters as in BK18: the amplitude, spatial and frequency spectral indices ($A$, $\alpha$ and $\beta$) for both the dust and synchrotron components, and a dust-synchrotron correlation parameter, $\epsilon$, derived from the BK18 likelihood. The dust emissivity index, $\beta_{d}$, is mildly correlated with $r$; other parameters have almost no correlation with $r$. The distribution for $r$ is also shown for foreground parameters fixed at best-fit values (red).} 
   \label{fig:r_fg_small}
\end{figure*}

\subsection{Foreground model}
\label{subsec:fg}
The BK18 foreground model assumes two polarized components: thermal dust and synchrotron emission, such that $D_\ell^{ij} =  D_\ell^{\rm{CMB}} + D_\ell^{{\rm FG},ij}$, and the foreground contribution is
\be
\label{eq:model}
D_\ell^{{\rm FG},ij} = D_\ell^{{\rm{dust}},ij}+  D_\ell^{{\rm{sync}},ij} + D_\ell^{{\rm{dust-sync}},ij}, 
\ee
for a cross-spectrum $D_\ell^{ij} \equiv \ell(\ell+1)C_\ell^{ij}/2\pi$ between maps $i$ and $j$, in thermodynamic units. The model assumes no polarized anomalous microwave or free-free emission, and that extragalactic contamination is negligible. The model terms are given by the following, written in simplified form appropriate for passbands that are delta functions in frequency; in practice the frequency scaling factors are computed by integrating across the telescope passbands.\\

\vspace{-0.1in}
\textbf{Dust} is assumed to follow a modified blackbody spectrum with temperature $T_d=19.6$~K. Its power spectrum model has three parameters: an amplitude, $A_d$, defined at $\nu_0=353$ GHz and $\ell_0=80$, a constant spectral index in multipole space, $\alpha_d$, and a constant emissivity index of the modified blackbody function in frequency space, $\beta_d$. The cross-power is given by
\be
D_\ell^{{\rm dust},ij}={A_d} \left(\frac{\ell}{\ell_0}\right)^{\alpha_d}
\left[\frac{\mu_d(\nu_i,\beta_d) \mu_d(\nu_j,\beta_d)}{\mu_d^2(\nu_0,\beta_d)}\right],
\ee
where $\mu_d(\nu,\beta_d)\equiv \nu^{\beta_d}B_{\nu}(T_d)g(\nu)$. Here $B_{\nu}(T_d)$ is the Planck function at frequency $\nu$, and the function $g(\nu) $ converts from flux to thermodynamic units.\\

\vspace{-0.1in}
\textbf{Synchrotron} is assumed to follow a power law as a function of both frequency, in antenna temperature, and multipole. Its power spectrum model also has three parameters: an amplitude $A_s$ defined at $\nu_0=23$~GHz and $\ell_0=80$, a spectral index in $\ell-$space, $\alpha_s$, and a frequency spectral index, $\beta_s$. The cross-spectrum is then
\be
D_\ell^{{\rm sync},ij}={A_s} \left(\frac{\ell}{\ell_0}\right)^{\alpha_s}
\left[\frac{\mu_s(\nu_i,\beta_s) \mu_s(\nu_j,\beta_s)}{\mu_s^2(\nu_0,\beta_s)}\right],
\ee
with frequency scaling $\mu_s(\nu,\beta_s)\equiv \nu^{\beta_s+2}g(\nu)$.\\

\textbf{Dust/synchrotron correlation} is modeled with an additional foreground parameter $\epsilon$, with the cross-spectrum given by
\ba
D_\ell^{{\rm dust-sync}, ij}&& = \epsilon \sqrt{A_dA_s} \left(\frac{\ell}{\ell_0}\right)^{\sqrt{\alpha_s \alpha_d}}\dot\nonumber\\
&&\left[\frac{\mu_d(\nu_i,\beta_d)\mu_s(\nu_j,\beta_s)+\mu_s(\nu_i,\beta_s)\mu_d(\nu_j,\beta_d)}{\mu_d(\nu_0,\beta_d)\mu_s(\nu_0,\beta_s)}\right]. \nonumber
\ea

\subsection{Likelihood}
The sky area covered by the BK18 data is small enough that the distributions of the bandpowers are non-Gaussian; the BK18 analysis uses the Hamimeche-Lewis approximation to model the non-Gaussian likelihood for the 66 spectra \citep{HamimecheLewis2008,Barkats2014}. In Appendix \ref{appendixA} we show the window functions, $w_{b\ell}$, that are used to bin a theory curve to compare to the data power spectra, to give bandpowers $D_b = \sum_\ell w_{b\ell} D_\ell$. These include the effect of transforming the apodization mask into Fourier space \citep{Knox1999,Barkats2014}.
The likelihood, $\mathcal{L}$, for the $BB$ model bandpowers, ${D}_b$, given the data bandpowers, $\hat{{D}}_b$, is given by
\begin{equation}
    -2 \log\mathcal{L}({D}_b|\hat{{D}}_b) = X_b \mathcal{M}_{bb'} X_{b'},
    \label{eq:HL}
\end{equation}
where the indices 
$b$ run over the 9 bands in $\ell$ and the 66 frequency auto- and cross-powers. 
$X_b$ is a vector of transformed $BB$ bandpowers, with expression given in Appendix \ref{appendixB}. The BK18 analysis uses signal and noise simulations to estimate the fiducial bandpower covariance matrix $\mathcal{M}_{bb'}$.

The theory model for the CMB, $D^{\rm CMB}_\ell$, is the sum of tensor perturbations with tensor-to-scalar ratio $r$ measured at $k=0.05$~Mpc$^{-1}$, with a scale-free spectral index $n_t=0$, and gravitationally lensed $B$-modes calculated using the \textit{Planck} 2018 best-fit $\Lambda$CDM parameters \citep{planck2018_parameters}. This is added to the foreground model, to give eight free parameters. These parameters all have uniform prior distributions, apart from the synchrotron index which has a Gaussian distribution with $-3.1\pm0.3$. The amplitudes $A_d$, $A_s$ and $r$ are required to be positive, the $\ell$-dependent slopes are limited to the range $-1<\alpha<0$, and the correlation coefficient is varied in the range $-1<\epsilon<1$.

\section{Tests of the foreground model}
\label{sec:tests}

The BK18 analysis shows the data are well fit by a three-parameter dust model plus the gravitationally lensed signal expected from \LCDM, with no evidence yet for non-zero synchrotron emission in the observed region, or for non-zero $r$. We initially reproduce these nominal constraints, shown in Figure \ref{fig:r_fg_small}. As noted in BK18, the foreground parameters now have little correlation with $r$, with only a mild correlation with the dust emissivity index, $\beta_d$. We highlight this by showing the distribution for $r$ where the foreground model parameters are fixed to their best-fit values; the upper limit on $r$ decreases by less than 10\% in this case, to $r<0.033$ at 95\% confidence. 

 An extensive suite of tests of the likelihood is reported in BK18, including internal consistency of splits of the data. Tests of the adopted foreground model have progressively increased with subsequent BICEP/Keck analyses. BK18 shows no significant impact on $r$ from imposing a dust index prior, including dust decorrelation as a function of wavelength, freeing the lensing amplitude, including $EE$ data, or dropping parts of the \textit{WMAP} or \textit{Planck} data. Here we explore some additional assumptions that can be tested at the likelihood level with the public products.

\subsection{Assumptions about the dust model}
Two elements of the model that might break down as the signal-to-noise increases are the assumption of a pure power law spectrum in $\ell$, and a spatially constant emissivity index. \\

\begin{figure}[!t]
  \centering
  \hspace{-0.3in}
  \includegraphics[width=1.1\linewidth]{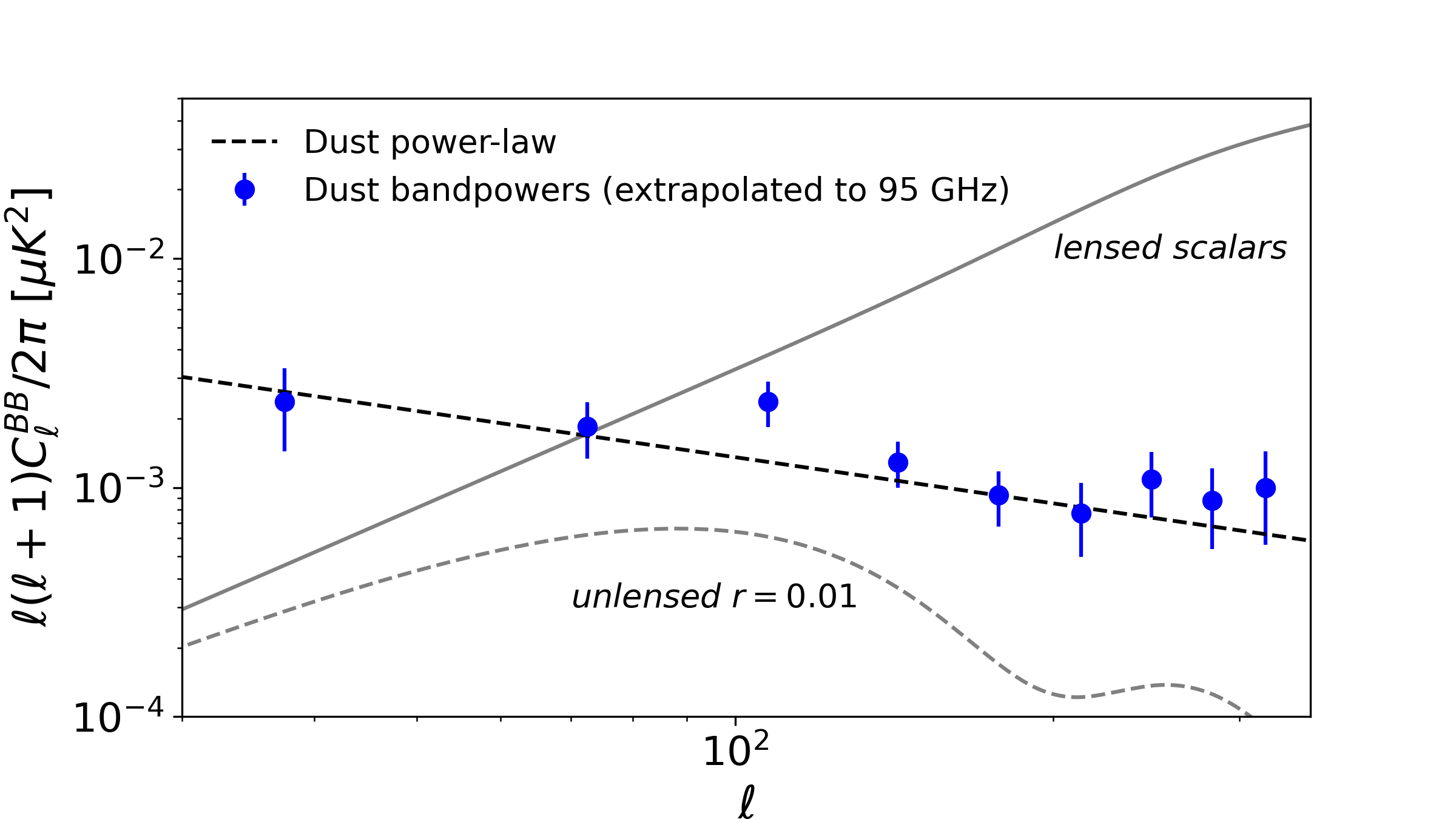}
  \includegraphics[width=1.025\linewidth]{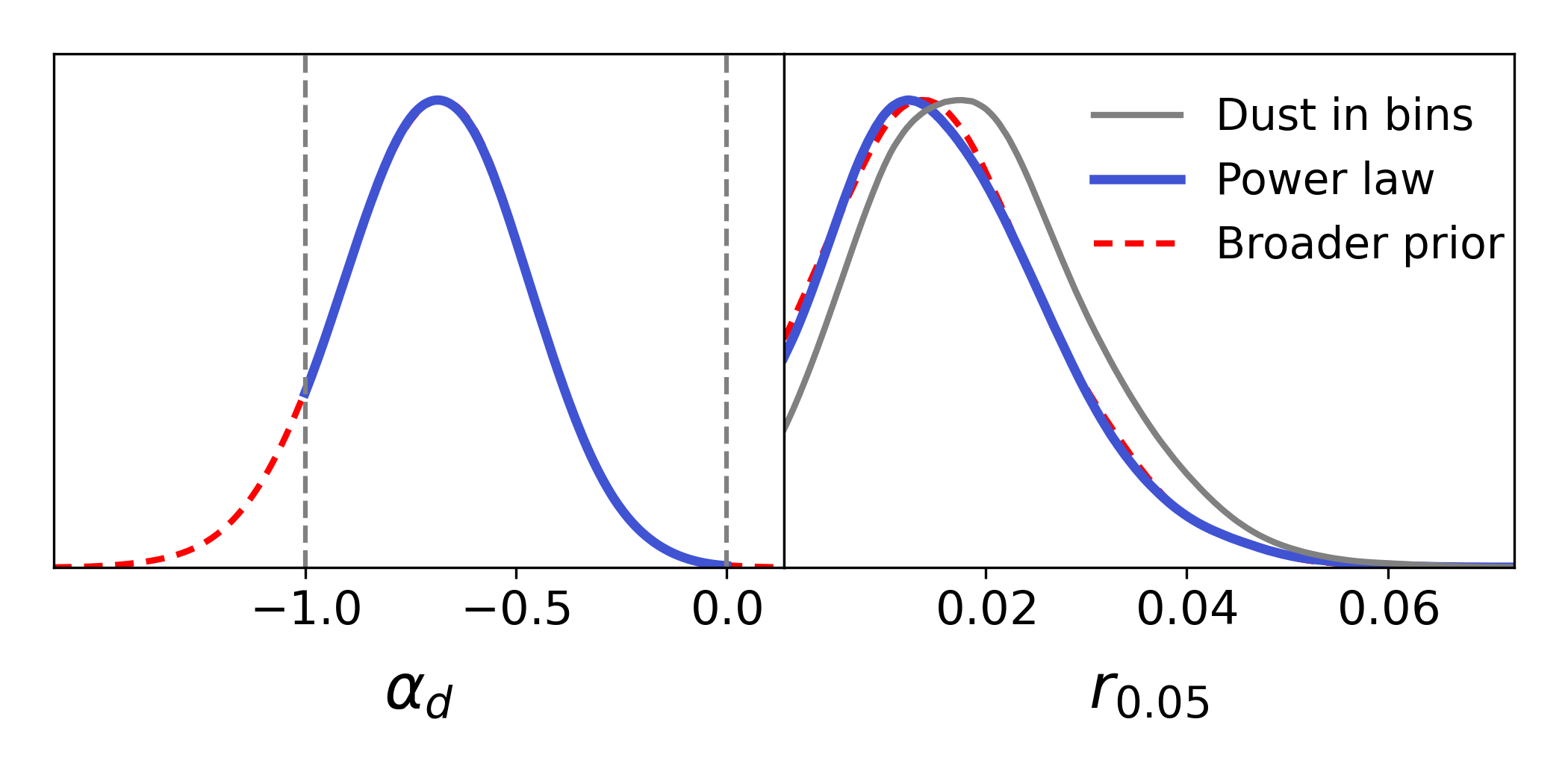}
  \caption{Top: the estimated dust power in multipole bins in the BK region, extrapolated to the 95~GHz bandpass, together with the best-fitting power-law model scaling as $(\ell/\ell_0)^{\alpha_d}$. Bottom left: the index, $\alpha_d$, can be constrained by the data without imposing a hard prior. Bottom right: we find a minimal impact on $r$ when the power-law model is expanded to have the per-bin amplitude, and negligible impact when the prior on the power-law exponent is relaxed. } 
   \label{fig:Bdust}
\end{figure}

{\bf Power-law spatial power spectrum}:
Observations by \planck\ show that dust follows a power-law spatially in the angular range $40<\ell<600$ \citep{planck2013_dust,planck2018_dust}. However, the \planck\ data show some departure from a power law in $BB$ at larger angular scales, and also show statistically significant variations in the power law index over different sky regions. Some departures from a power law, and a detection of a spatially varying slope, are also reported in \cite{cordova/etal:2023} with ACT and WISE data. As data improve, we might see a deviation from a pure power-law in the BICEP/Keck region. To explore this, we extend the BK18 likelihood to include a different amplitude of dust in each of the nine bands, following a similar approach as used in \citet{wolz2023} for simulated Simons Observatory data. We replace the two parameters $A_d$ and $\alpha_d$, where $D_\ell = A_d(\ell/\ell_0)^{\alpha_d}$ at 353~GHz, with nine dust amplitudes that are sampled simultaneously with $r$ and the other foreground parameters (making 15 parameters altogether instead of eight). We find a minimal impact on $r$, as shown in Figure~\ref{fig:Bdust}. We also find no significant departure from power law for the dust, although observe that the dust amplitude fluctuates about 2$\sigma$ higher than the model in the $\ell \sim 100$ bin. The dust amplitudes are shown in Figure~\ref{fig:Bdust}, extrapolated to the BICEP3 95~GHz passband. They are consistent with estimates in BK18's Figure 16 where the dust and non-dust parameters are estimated separately in each bin, but with smaller uncertainties since here we fit a common model simultaneously to the nine bins.  

Within the power-law model, the BK18 analysis imposes a hard prior $-1<\alpha_d<0$, motivated by \planck\ data. Given the improved data quality in BK18, we expand the prior on the dust slope, $\alpha_d$, so that it is not limited by the hard boundaries. The BK18 data can now constrain this slope, as shown in Figure~\ref{fig:Bdust}. Loosening this prior has a negligible impact on $r$. \\

\begin{figure}[!bth]
  \centering
 \includegraphics[width=\linewidth]{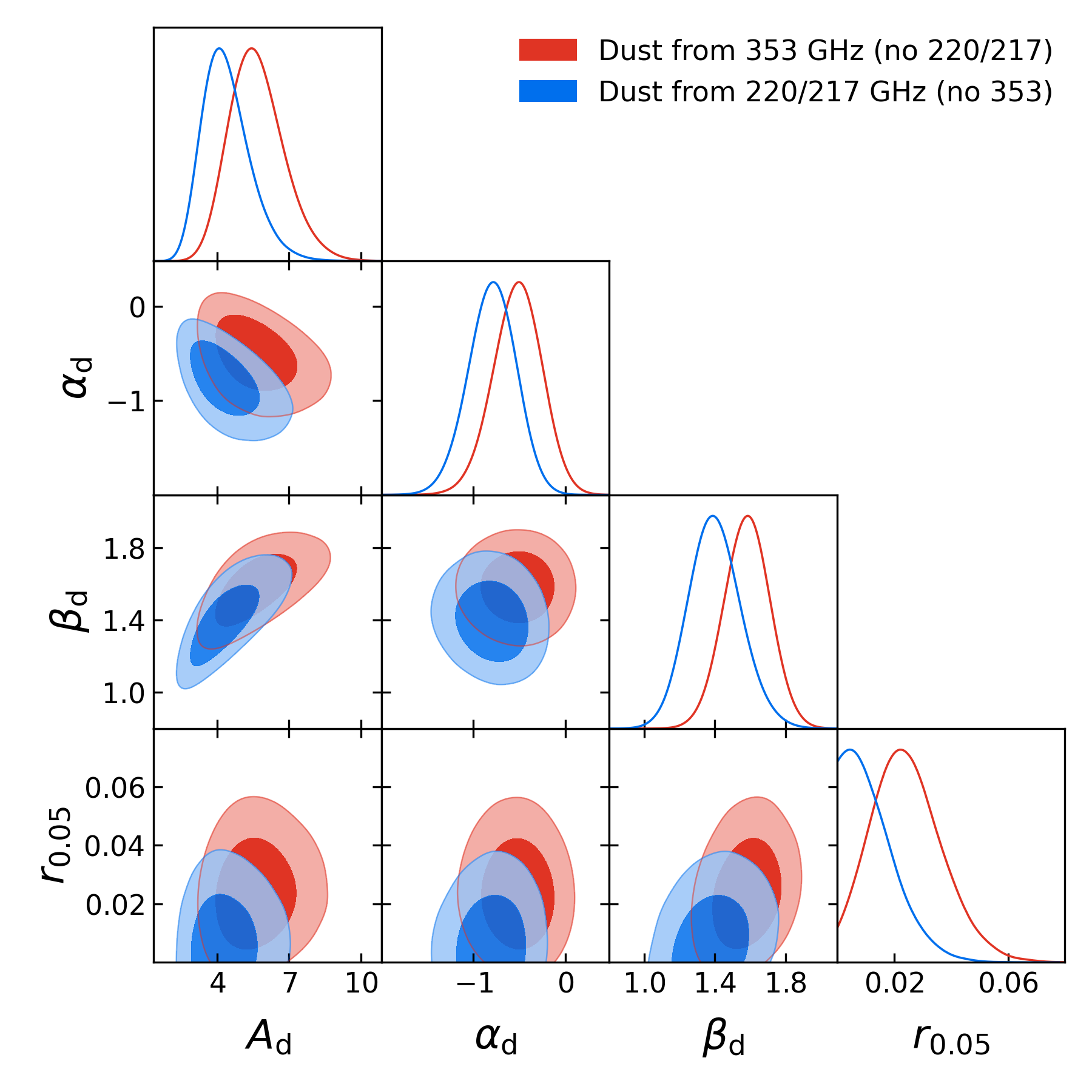}
  \caption{Parameter estimates for $r$ and the dust amplitude and indices are consistent whether 220+217 GHz or 353~GHz data are used to clean the dust, with a typical variation in the mean values of $\sim1\sigma$. Scatter is expected since the data measuring the dust are independent in the two cases.}
   \label{fig:220_353}
\end{figure}

{\bf Constant index:} If the model for a spatially constant dust emissivity index is correct, the maps of the dust at different frequencies will be perfectly correlated. This was tested in the BK18 analysis by varying a decorrelation parameter, found to be consistent with zero. This variation can also be captured with a moment expansion of the index, as described in \cite{azzoni/etal:2021,wolz2023}.
 In \cite{azzoni/etal:2021}, minimal impact on the earlier BK15 $r$ limit was found when a moment expansion was included, similar to the results including the decorrelation parameter. Another simple test is to show consistency of $r$, and the emissivity index, cleaned with dust maps measured at different frequencies. In Figure~\ref{fig:220_353} we show $r$ and the three dust parameters for the case where both the 217 and 220~GHz data from BICEP2/Keck and \planck\ are discarded, and 353~GHz is used as the main dust tracer (`Dust from 353'), versus discarding 353~GHz and using 217/220~GHz as the main dust tracer (`Dust from 220/217')\footnote{The latter `Dust from 220/217' case, discarding 353~GHz, was shown in BK18's Figure 21.}. In both cases the usual synchrotron parameters are also sampled. We find that the estimated model parameters are statistically consistent, with parameter means differing by $\sim1\sigma$, but still supporting the spatially-constant index assumption. The best-fit value of $r$ is non-zero in the `Dust from 353' case, but there is no evidence for a detection. In both cases the dust parameters can be constrained to similar precision. \\

\begin{figure}[!bt]
  \centering
 \includegraphics[width=0.515\linewidth]{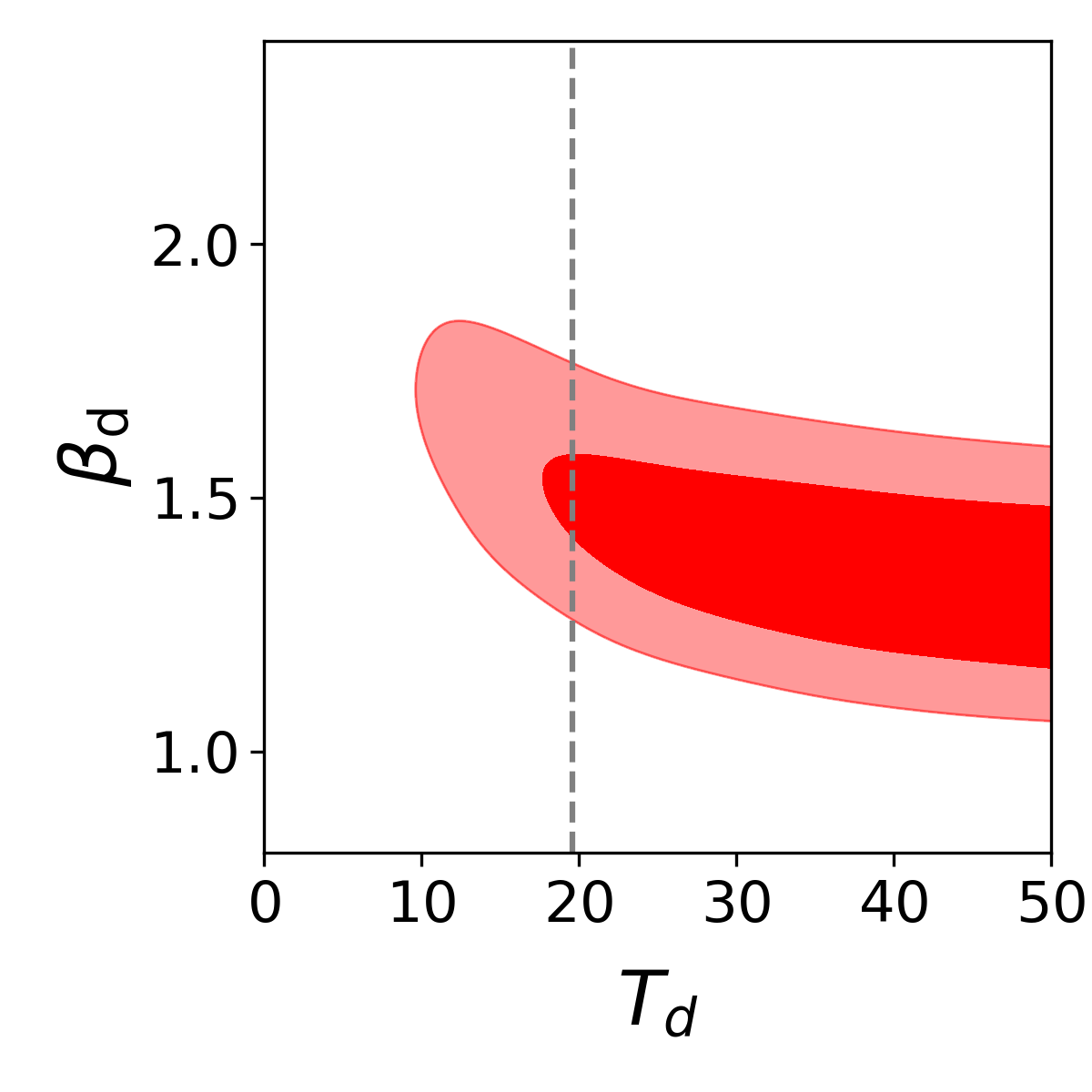}
 \includegraphics[width=0.47\linewidth]{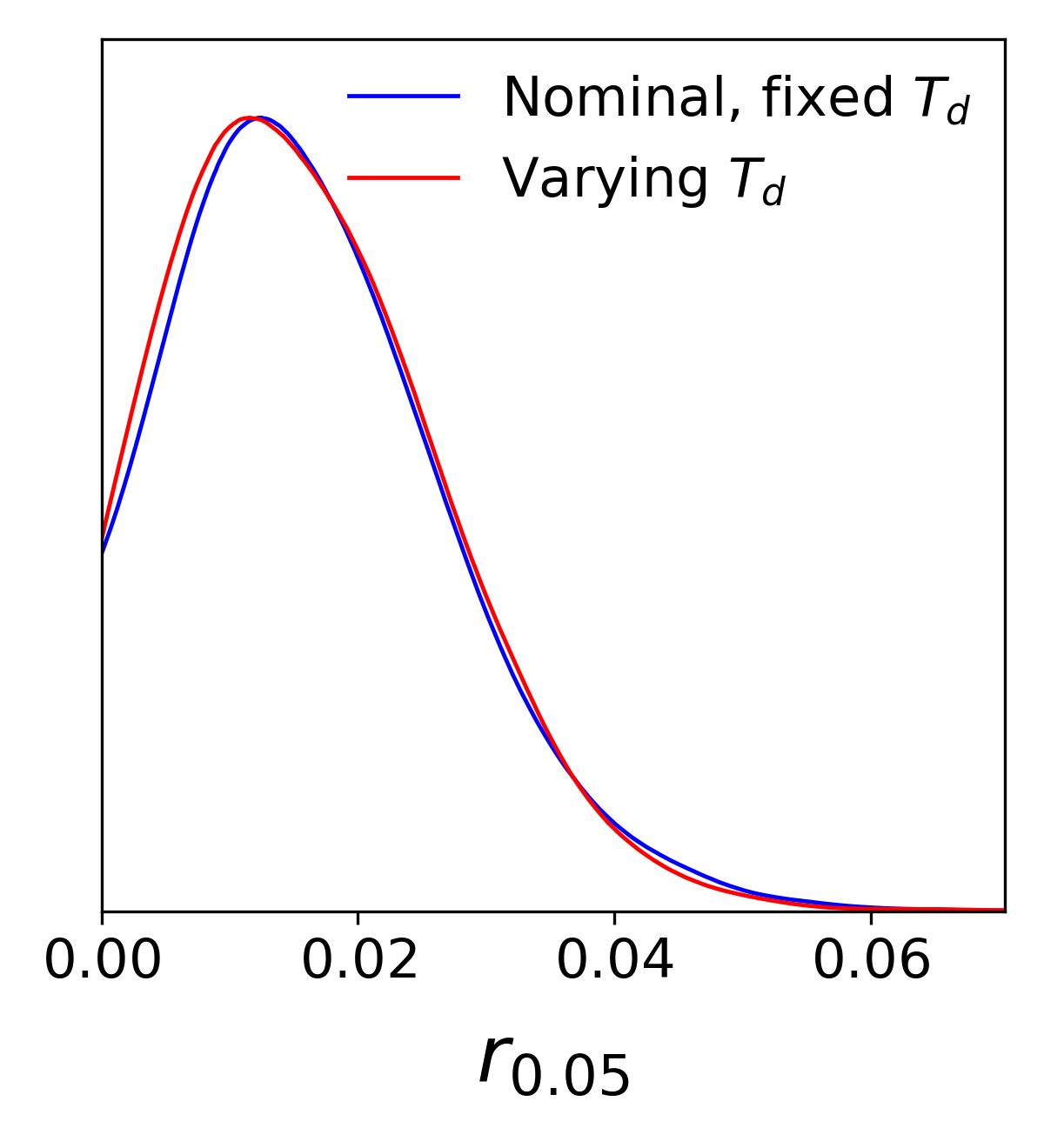}    
  \caption{Varying the dust temperature, $T_d$, does not impact the constraint on $r$. The temperature cannot be bounded from above by this dataset, and is weakly anti-correlated with the dust emissivity index, $\beta_d$.}
   \label{fig:Tdust}
\end{figure}

{\bf Fixed dust temperature:} The fiducial BK18 model assumes a dust temperature of $19.6$~K, which is the mean temperature estimated by \planck\ over the sky. Although the temperature may depart from the mean value in the particular BK18 region, we would not expect the BK18 data to be sensitive to the choice since the highest frequency is far from the peak of the modified blackbody distribution. We confirm this in Figure~\ref{fig:Tdust}, showing the negligible effect on $r$ if the temperature is varied over a broad range, $0<T_d<50$~K.

\begin{figure}[bt]
  \centering
 \hspace{-1.1in}
\includegraphics[width=0.95\linewidth]{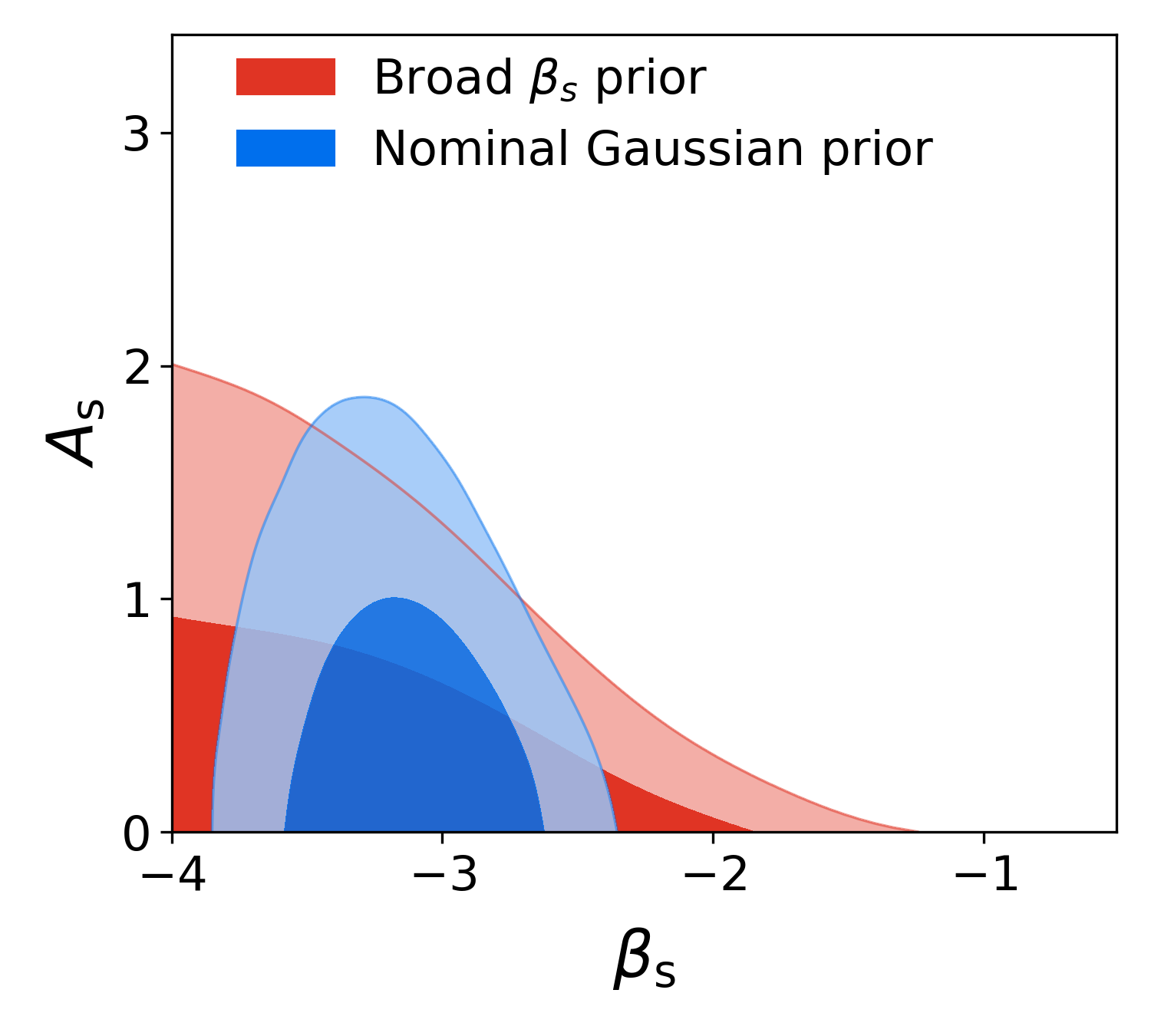}
\hspace{-1in}

\vspace{-2.3in}
\hspace{1.5in}
\includegraphics[width=0.4\linewidth] {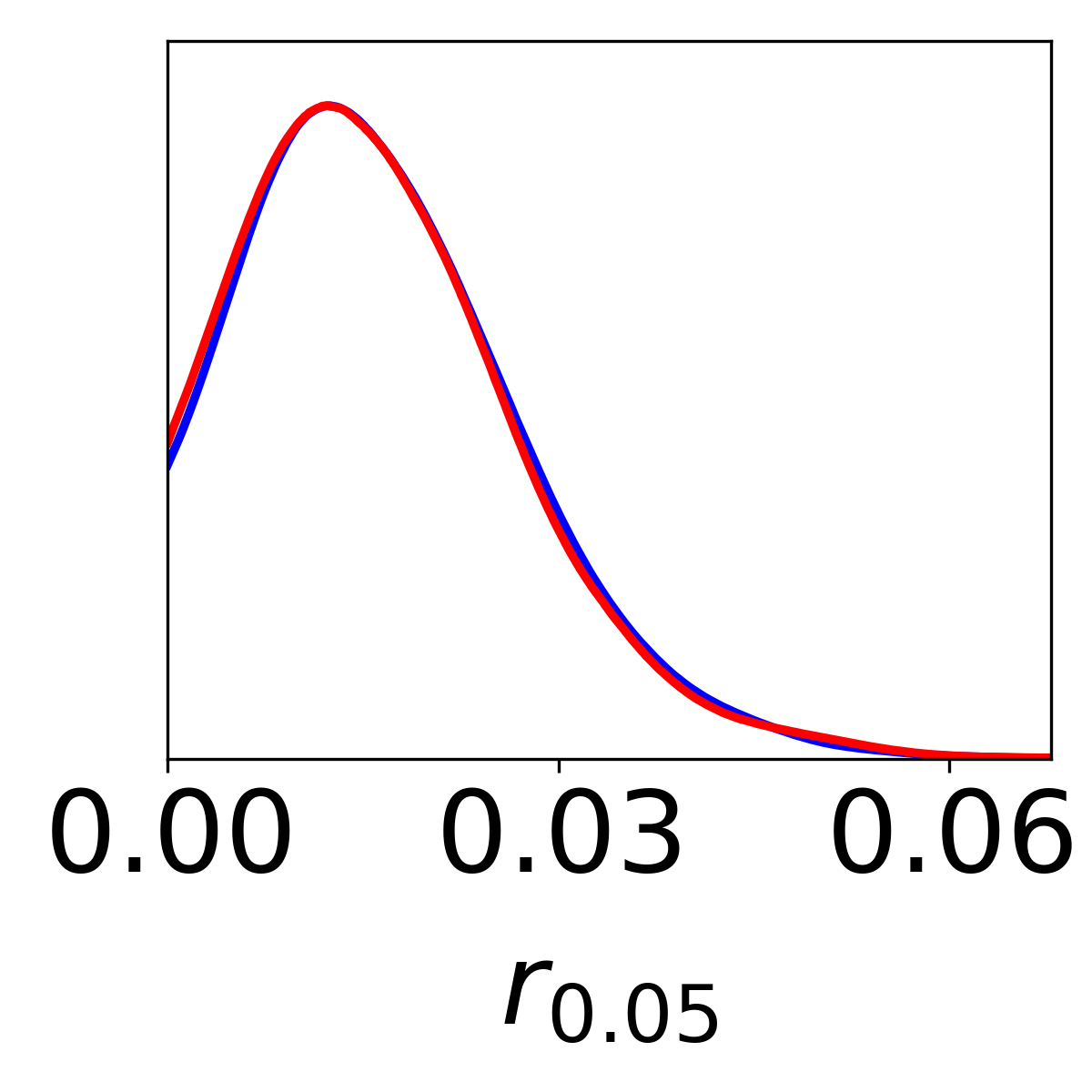}

 \vspace{1in}

 \vspace{-0.2in}
  \caption{Inset: removing the Gaussian prior on the synchrotron index has a negligible effect on $r$. Main: even though no synchrotron emission is detected in the BK18 data, more negative values of $\beta_s$ are preferred when a uniform prior is imposed on the index, given the larger volume of possible models, and $\beta_s>-2$ is disfavored.}
   \label{fig:sync_index}
\end{figure}

\subsection{Assumptions about synchrotron}

Since no synchrotron is detected in BB in this region, the only modeling assumption we consider checking is to {\bf relax the prior on the synchrotron index.} In the nominal model there is a Gaussian prior on the synchrotron index, with $\beta_s=-3.1\pm0.3$ estimated from \wmap\ data. The upper limit on the synchrotron amplitude is $A_s<1.5~\mu$K$^2$ at 95\% CL, defined as the power at 23~GHz and $\ell=80$.

The data can still be used to put an upper limit on the synchrotron index, assuming a flat prior, since although a flat index $\beta_s=0$ should be able to fit the data as $A_s$ tends to zero, there is a smaller volume of models at this high-$\beta$ limit. Figure~\ref{fig:sync_index} shows that there is minimal impact on $r$ if the synchrotron index is varied uniformly within a prior range of $\-4<\beta_s<0$.

\subsection{Unknowns: consistency of 95~GHz and 150~GHz}
\begin{figure}[!t]
  \centering
   \includegraphics[width=\linewidth]{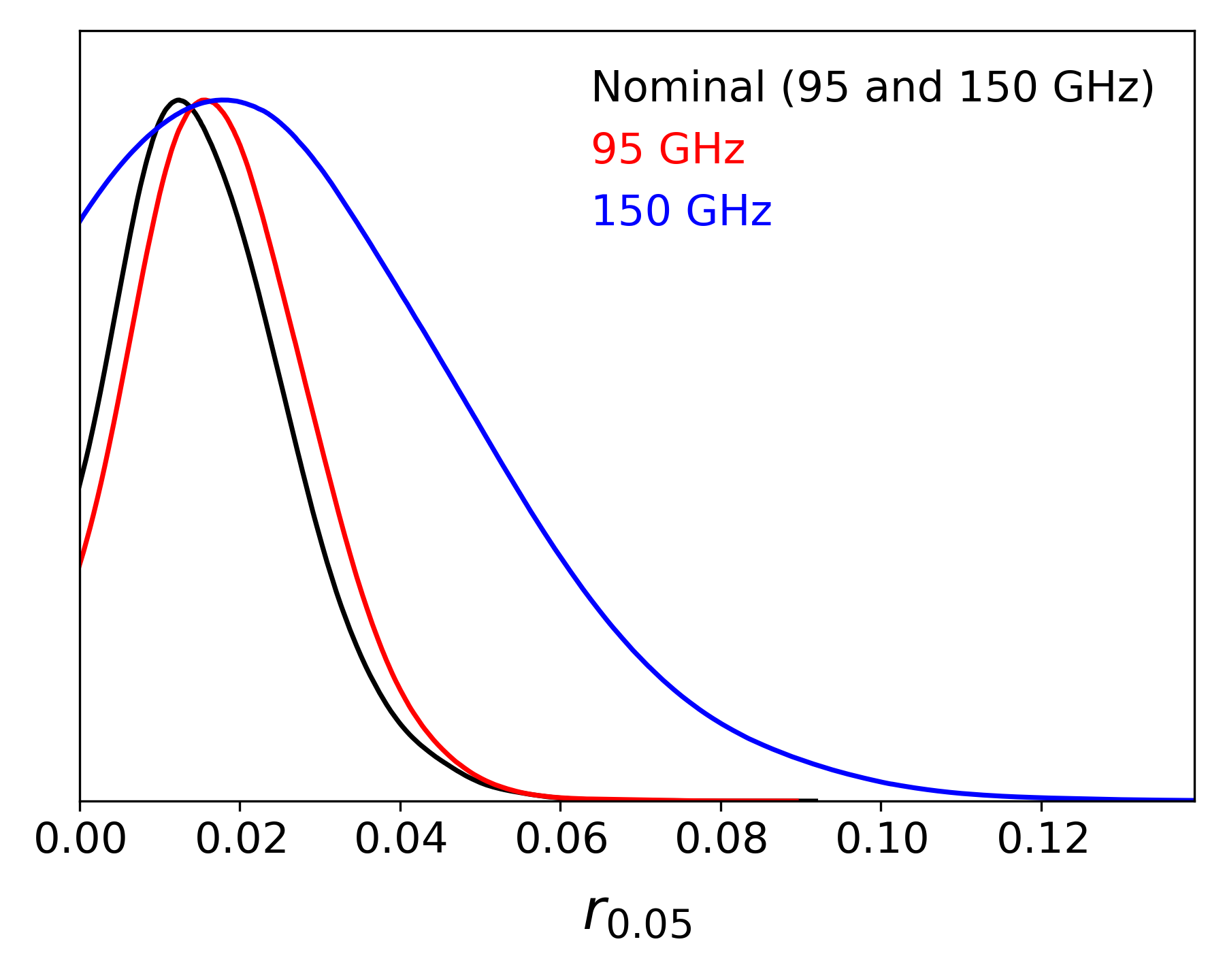}
  \caption{Distribution for $r$ estimated from 95~GHz or 150~GHz, compared to the nominal combination. The foreground frequencies are included in all cases. The current limit on $r$ is dominated by the 95~GHz data.}
   \label{fig:95_150}
\end{figure}

A useful consistency test is to show agreement of the $r$ measurement from different frequencies, to test for unexpected modeling errors (from non-zero polarized AME, magnetic dust, etc). Figure~\ref{fig:95_150} shows the distribution for $r$ estimated from just the BK18 95~GHz data, or just the 150~GHz data, in each case including the ancillary frequency data used to clean foregrounds (23--44 and 217--353~GHz). The results are consistent, but as noted in BK18, the latest BICEP3 95~GHz data dominate the constraint, with the overall $r<0.036$ upper limit driven by the 95~GHz data: the constraint is $r<0.038$ without the 150~GHz data. The 150~GHz data, in combination with the foreground tracers, give $r<0.072$ at 95\% CL; this larger uncertainty sets the current limit on a frequency null test.

\section{`BK-lite' - a compressed likelihood}
\label{sec:compress}

\begin{figure*}[!bth]
  \centering
\includegraphics[width=0.9\linewidth]{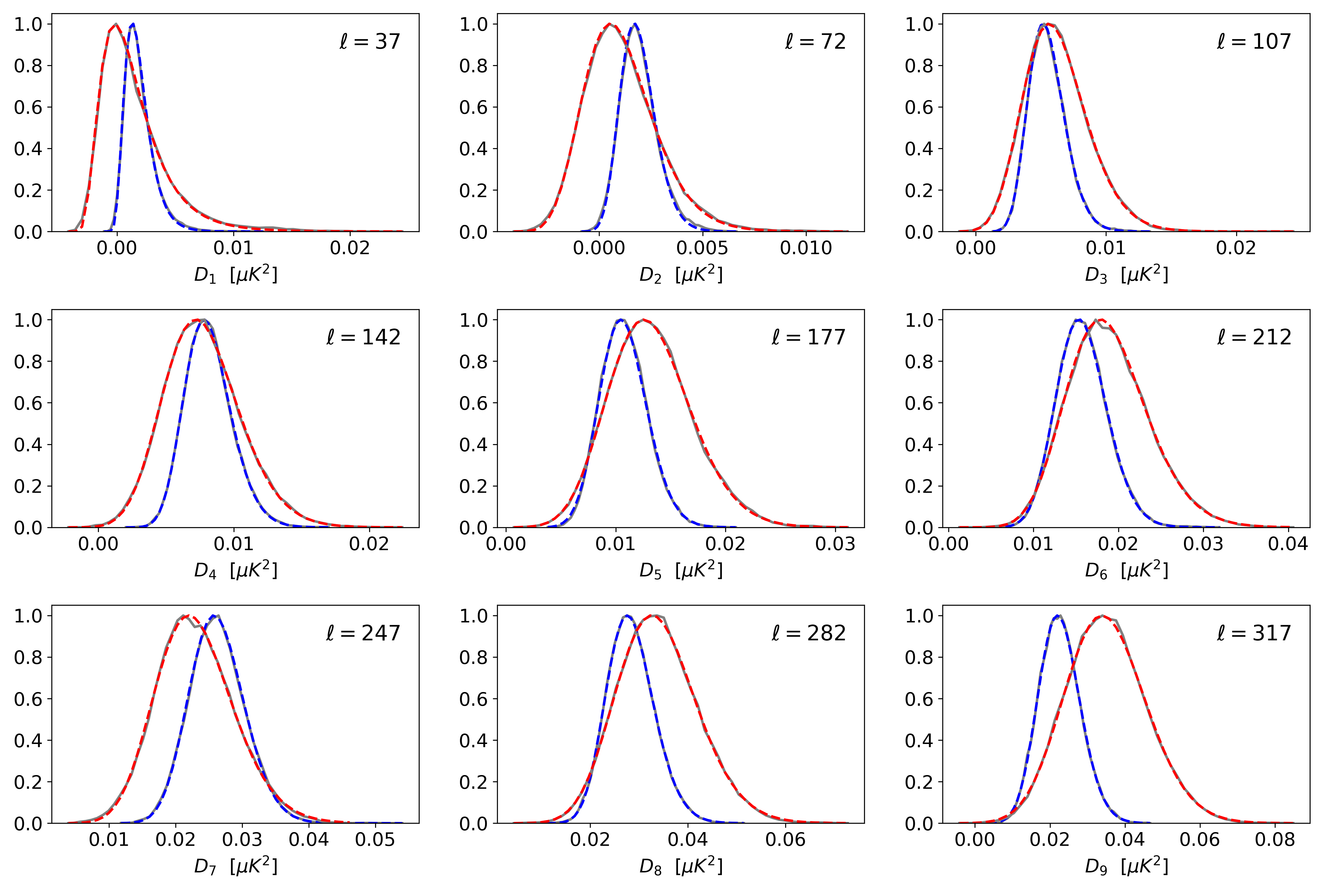}
  \vspace{-0.2in}
    \caption{Probability distributions for the nine CMB bandpowers estimated simultaneously from data in the BICEP3 sky region (blue) and in the smaller BICEP2/Keck region (red), with a common foreground model. Each distribution is fit with a three-parameter offset log-normal distribution (gray). 
    }
   \label{fig:lognormals_e}
\end{figure*}

In this section we construct a foreground-marginalized likelihood, by estimating the blackbody CMB power in each of the nine $\ell$-bandpowers. In doing so we keep the same foregound model as used in BK18, motivated by the tests of the previous section and those reported in BK18. This marginalization approach has been applied to ACT and SPT data, and used for the Plik\_lite \textit{Planck} likelihood~\citep{Dunkley2013, Calabrese2013, planck2015_likelihoods,choi/etal:2020}, and has been implemented for one of the Simons Observatory likelihood pipelines \citep{wolz2023}.

Since the BK18 data have different bandpower window functions for the BICEP3 versus BICEP2/Keck spectra due to the different sky coverage, shown in the Appendix, there is not a single set of uniquely defined bandpowers to estimate. Instead, one could estimate nine bandpowers for each of the window function shapes, i.e. a vector of $9\times3$ bandpowers if we approximate the suite of window functions as having the same shape for each frequency in each region (or cross-region). This approach was used in the ACT analyses \citep{Dunkley2013,choi/etal:2020}, which included spectra estimated from different sky regions.

Here we demonstrate the method on two subsets of the BK data. First we use only the data in the BICEP3 region, which includes the BICEP3 95 GHz map and the \textit{WMAP} and \textit{Planck} maps, all of which have a common window function. The constraint on $r$ derived from just these eight maps (with $r<0.037$) is similar to the constraint from the full 11 maps, as already shown in Appendix E2 and Figure 20 of BK18, and reproduced here in Appendix \ref{appendixC}.  

In a second case we use data from all 11 maps, but only include auto and cross-spectra computed from maps that use a common window function, i.e. either in the BICEP3 region, or in the smaller BICEP2/Keck region. We exclude cross-spectra between map pairs from both sky regions, such as Keck-220~$\times$~Planck-353. For the data in the smaller region, we also revert to using the BK15 data for demonstrating this method \cite{BK15}, since these have normalized window functions in the publicly available likelihood code\footnote{The window functions in BK18 are not normalized and have different amplitudes for the 95, 150 and 220 GHz spectra.}. This combination of data results in $r<0.032$ at 95\% CL, and the estimated parameters are also shown in Appendix \ref{appendixC}. 

\begin{figure*}[!th]
  \centering
 \vspace{-0.3in}\includegraphics[width=0.8\linewidth]{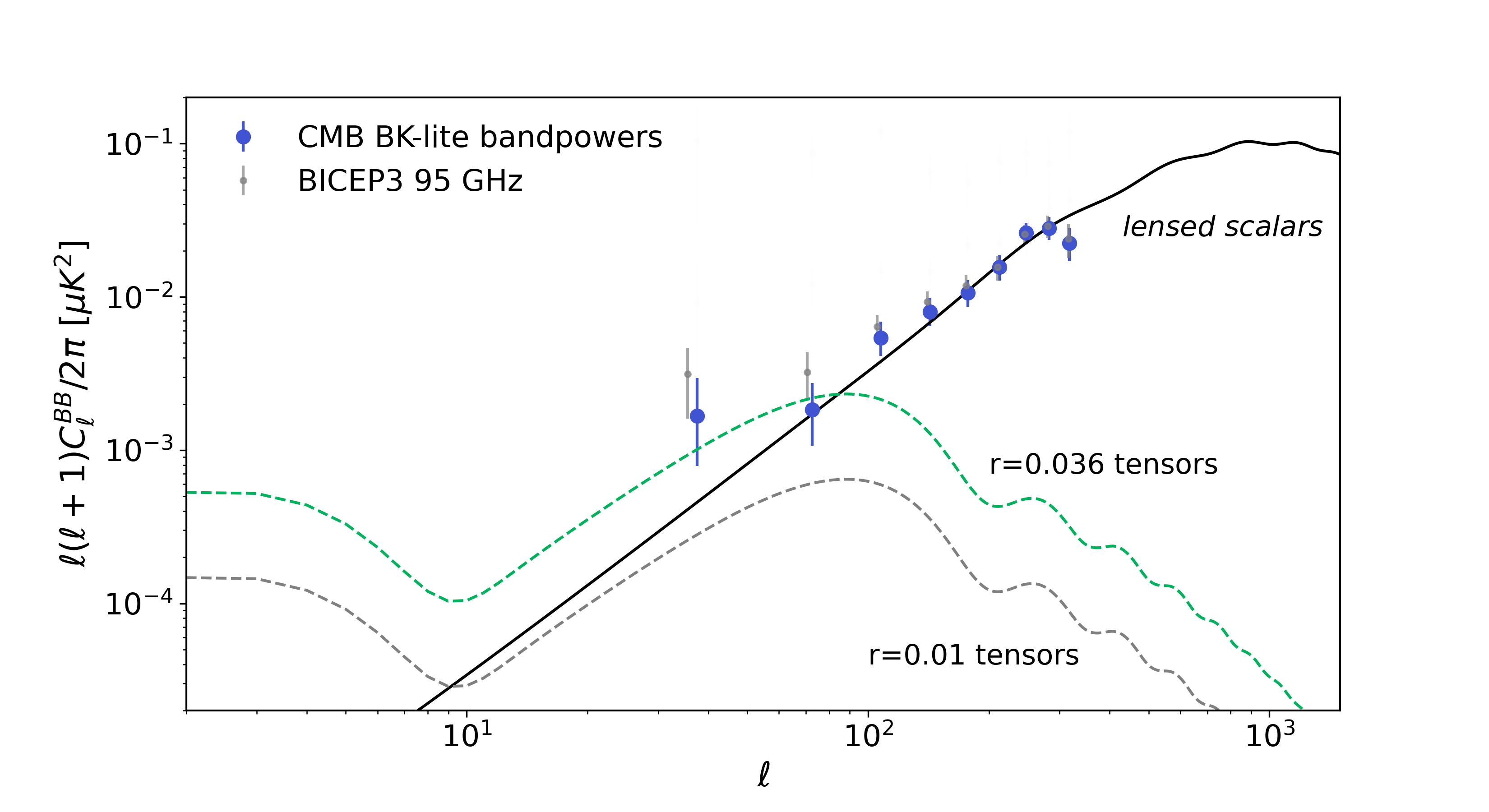}
 \vspace{-0.15in}
  \caption{Foreground-marginalized CMB $BB$ bandpowers derived from BICEP3, {\it Planck} and {\it WMAP} data on the BICEP3 sky region. The bandpower values are the median of the best-fitting log-normal distribution, and the errors indicate the 16th and 84th percentiles of the distribution. Theoretical $B$-mode power spectra are shown for the lensed scalars, and for tensor contributions for $r=0.036$ (current upper limit) and $r=0.01$ (near-term target), for a $\Lambda$CDM model.
  }
   \label{fig:cmb_marg_power}
\end{figure*}

\subsection{Bandpower estimation}

For the first case we estimate the CMB bandpowers, $D_b^{\rm CMB}$, in the nine $\ell$ bins simultaneously, marginalizing over the seven foreground parameters. Here the total model bandpowers are given by 
\begin{equation}
D_b^{ij} =D_b^{\rm CMB} + \sum_\ell w_{b\ell} D_\ell^{{\rm FG},ij},
\end{equation}
where the window functions correspond to those for the BICEP3 region. In practice we implement this by modifying the {\texttt{Cobaya}} code to sample each bandpower as a separate parameter instead of sampling the tensor-to-scalar ratio $r$. The same seven foreground parameters are included to compute $D_\ell^{\rm{FG},ij}$, and we sample the 16 parameters using the Metropolis-Hastings algorithm. 
For the second case we estimate $9\times2$ CMB bandpowers -- nine to fit the spectra computed from the BICEP3 region, and nine for the spectra from the BICEP2/Keck region --  and the same seven foreground parameters. We find the estimated foreground parameters are consistent when estimating the bandpowers or when directly constraining $r$, shown in Appendix \ref{appendixC}. 

 We show the 1D posterior distributions for the BK18 CMB bandpowers in Figure~\ref{fig:lognormals_e}, for the second case.
As expected, the distributions are non-Gaussian. This was not the case for the ACT and \planck\ CMB bandpower estimates, nor is expected to be for Simons Observatory \citep{wolz2023}, since those surveys cover smaller angular scales and/or larger areas, where the distributions are Gaussian to good approximation.  
Following e.g., \cite{bond/jaffe/knox:2000}, we fit the one-dimensional distributions with offset log-normal distributions of the form
\be
p(D_{b})= \frac{1}{(D_{b}-D_0) \sigma \sqrt{2\pi}} e^{-(\ln (D_{b}-D_0)-\mu)^2/(2\sigma^2)},
\label{eq:ch4_lognorm}
\ee
where $D_0$ is the offset that makes $\ln (D_{b}-D_0)$ normally distributed for each bin, and $\mu$ and $\sigma$ are the mean and standard deviation of $\ln (D_{b}-D_0)$ respectively.
These best-fitting log-normal distributions are also shown in Figure~\ref{fig:lognormals_e}. 
We estimate the covariance of the $\ln (D_{b}-D_0)$ parameters from the MCMC chains, derived using the best-fitting $D_0$ for each bin.
We find correlations of up to 20\% between neighbouring bandpowers in $\ell$, and between bandpowers that cover the same $\ell$ range for the two window functions. Examples of the 2D distributions are shown in Appendix \ref{appendixC}.

The estimated bandpowers are shown in Figure~\ref{fig:cmb_marg_power} for the BICEP3 region in the second case, indicating the median as the data point, and the 16th and 84th percentiles of each one-dimensional distribution as the errors.  
Our method of simultaneously estimating the nine bandpowers while marginalizing over the foregrounds is different to BK18's Figure 16, 
where for plotting purposes each bin is decomposed independently into CMB, dust, and an upper limit on synchrotron. Here data from all the bins are used to constrain the foregrounds simultaneously. 
On degree-scales and larger ($\ell < 200$) the CMB bandpowers are lower than the total 95 GHz spectrum because the dust contribution has been removed, while at smaller scales the impact of dust removal is negligible at this frequency.

\subsection{A BK-lite likelihood}

Given the correlation observed between bins, we use a covariant lognormal likelihood 
with $\ln \mathcal{L}$ given by
\ba
 -\frac{1}{2}  \left(\ln(\mathbf{D}_{\rm b}-\mathbf{D_0})-\boldsymbol{\mu}\right)^T \mathbf{Q}^{-1} \left(\ln(\mathbf{D}_{\rm b}-\mathbf{D_0})-\boldsymbol{\mu}\right) \nonumber \\ 
- \sum_{i=1}^{\rm nbin} \ln(D_{\rm b, i}-D_{0,i}),
\label{eq:indep_LN_like}
\ea
where $\mathbf{D}_{\rm b}$ is the vector of the binned theory spectrum and $\mathbf{D_0}$ and $\boldsymbol{\mu}$ are vectors of the best-fit offsets and means for the CMB data lognormal bandpowers.
The binned theory spectrum in our first case is binned using the BICEP3 bandpower window functions; the second case concatenates a theory spectrum binned with the BICEP3 window functions, with one binned with the BICEP2/Keck window functions. To reduce the impact of noise in the covariance matrix estimation, we neglect covariance between bins where the correlation coefficient is estimated to be $<5\%$. 
\begin{figure}
  \centering
   \includegraphics[width=\linewidth]
   {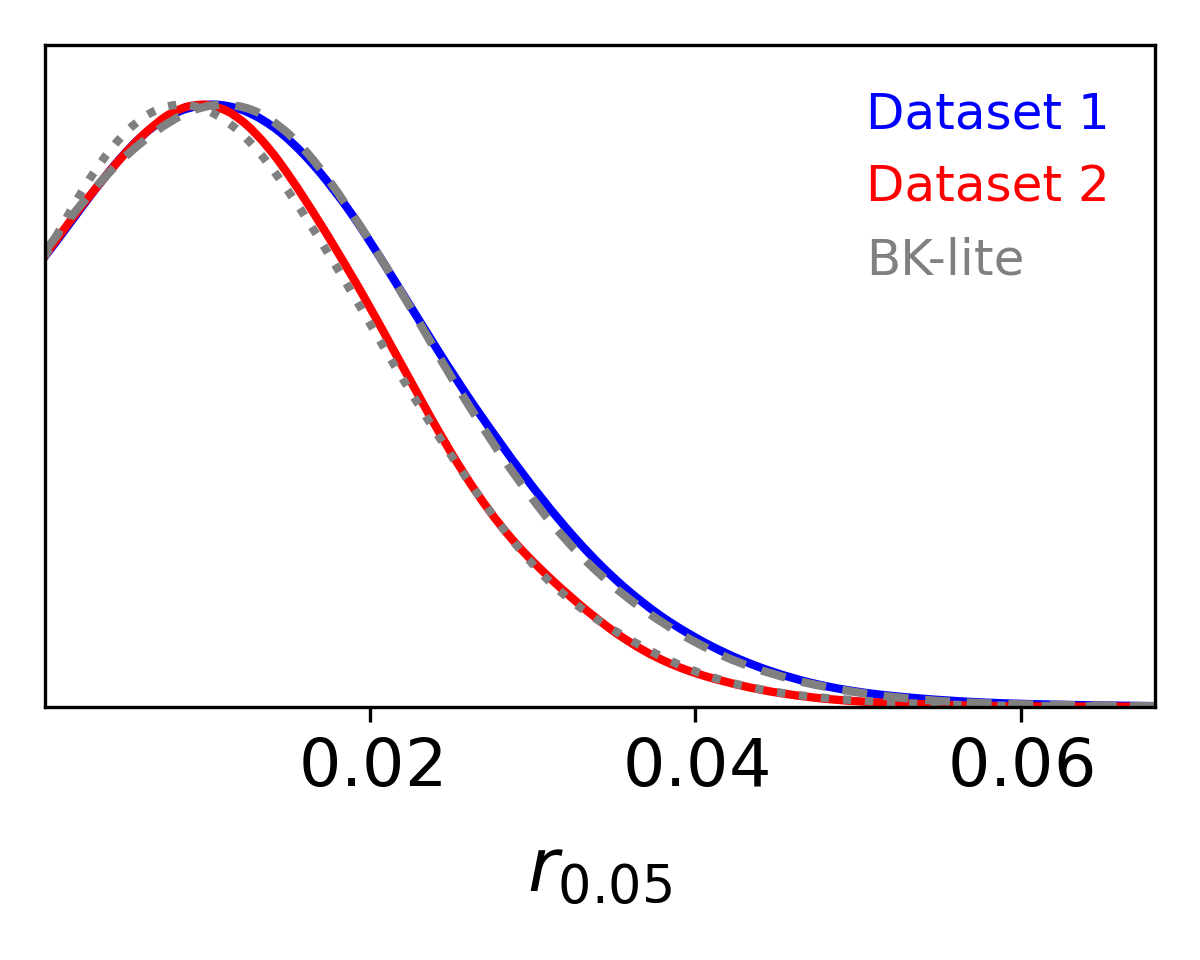}
 \vspace{-0.35in}
  \caption{The constraints on $r$ from our foreground-marginalized `BK-lite' likelihood agree with the nominal BK18 likelihood, restricted to the spectra from maps that use the BICEP3 footprint (`Dataset 1') or spectra from all the maps but discarding cross-spectra between maps with different footprints (`Dataset 2').}
   \label{fig:1D_r}
\end{figure}

The constraints on the tensor-to-scalar ratio $r$ using our BK-lite likelihood, with only one free parameter $r$, and the full BK likelihoods with eight free parameters, are shown in Figure~\ref{fig:1D_r} for the two subsets of data. The distributions for $r$ agree well, giving the same 95\% upper limit in both cases. If the neighboring-bin correlation is neglected in the marginalized likelihood, we find a small shrinkage of the distribution. The distribution for $r$ also agrees in the case where \LCDM\ parameters are allowed to vary and the \planck\ data are included.

To extend this to the full likelihood one could include a third vector of bandpowers for the `cross-window' spectra (i.e. for BICEP3 x BICEP2 spectra). This is straightforward in principle, but implementing it in the available likelihood code would be more complicated than this current implementation, so we leave it for a future exercise.

\section{Conclusion}
\label{sec:discussion}

In this paper we have re-examined the likelihood analysis for the BICEP/Keck collaboration's constraint on the tensor-to-scalar ratio, and further explored effects of different foreground modeling choices on the results.
We then estimated the joint probability distribution of nine foreground-marginalized CMB bandpowers. 
We fit these bandpowers with offset log-normal distributions and constructed a `BK-lite' likelihood that reproduces the multi-frequency likelihood when using the same data products. A likelihood of this form is useful because it does not require re-sampling the foreground parameters for each new estimate of $r$ in joint-probe analysis. The CMB-only bandpowers can also be visually compared to the CMB theory power spectrum, and it can be easily used with automatically differentiable theory codes when using sampling methods such as Hamiltonian Monte Carlo. A Gaussian implementation of this method has already been developed for the Simons Observatory, but the log-normal likelihood used here can be directly applicable to the South Pole Observatory, CMB-S4, and other experiments which target small sky areas.

\section*{Acknowledgements}

We thank David Alonso for useful discussion. We acknowledge discussions with members of the Simons Observatory Analysis Working Groups. We are grateful to the BICEP/Keck team for making their likelihood publicly available. We used the {\texttt{Cobaya}} and {\texttt{GetDist}} codes.
EC acknowledges support from the European Research Council (ERC) under the European Union’s Horizon 2020 research and innovation programme (Grant agreement No. 849169).

\bibliography{bicep}

\appendix

\section{Window Functions}
 \label{appendixA}
 \begin{figure}
\vspace{-0.3in}

  \centering
\includegraphics[width=1.1\linewidth]{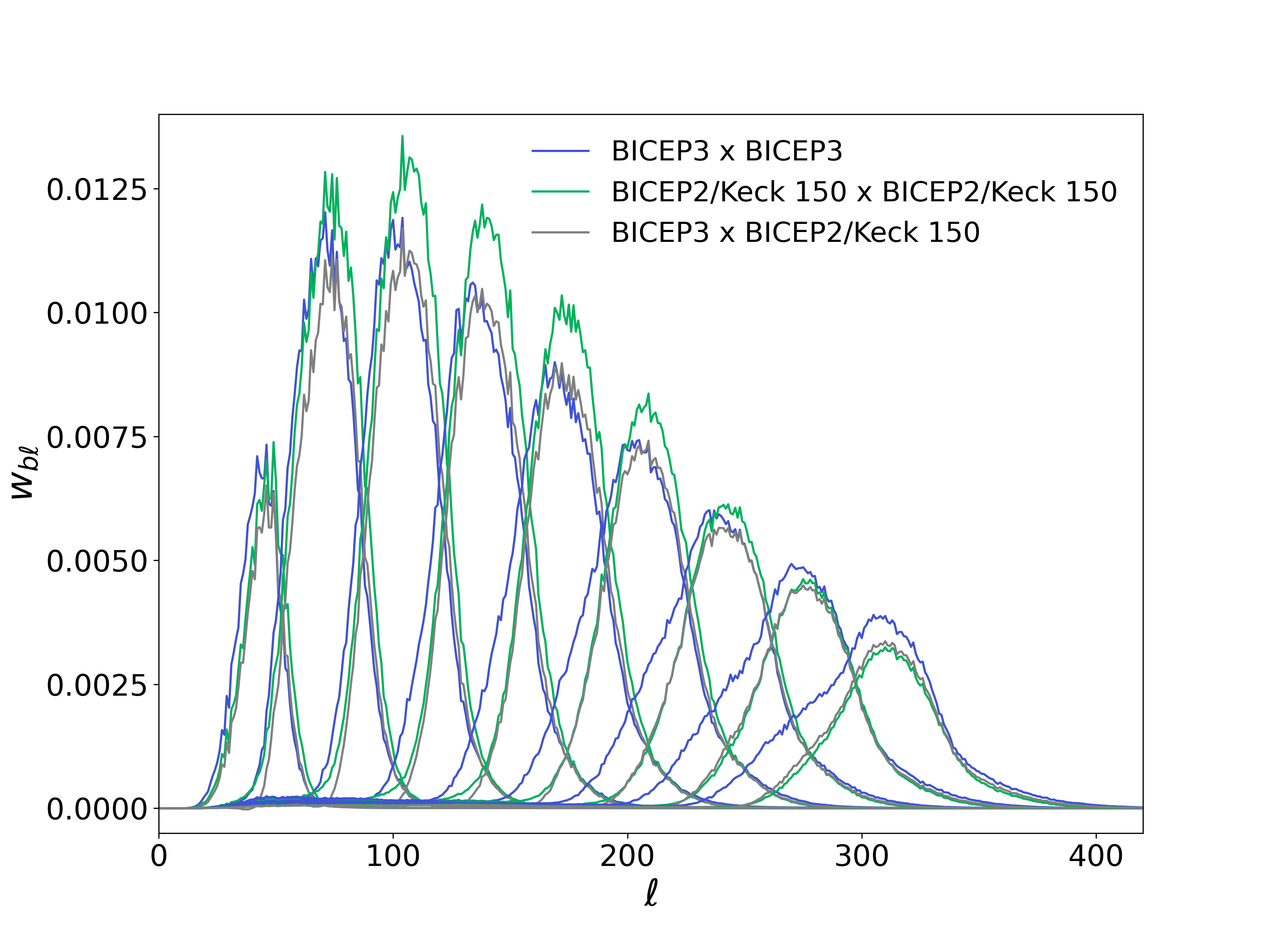}
\vspace{-0.3in}
  \caption{The un-normalized BK18 bandpower window functions for the 9 bins, part of the BK18 public data products. They differ for spectra estimated in the larger BICEP3 region (blue) versus the smaller BICEP2/Keck region (green). Cross-spectra between the two regions have a third typical window function shape (gray).}
   \label{fig:windows}
\end{figure}
The bandpower window functions are defined such that $D_b = \sum_\ell w_{b\ell} D_\ell$. There are effectively three distinct sets of window function shapes: one for the larger BICEP3 field which applies to the BICEP3, \textit{WMAP} and \textit{Planck} auto-spectra, one for the smaller BICEP2/Keck field which applies to the BICEP2/Keck auto-spectra, and one set for cross spectra between the two different fields. These un-normalized window functions are shown in Figure~\ref{fig:windows} for a sample cross-spectrum. The BICEP2/Keck window functions have different normalizations at the three different frequencies, although the shapes are almost the same. The cross spectra between BICEP3 regions and BICEP2/Keck regions have different normalizations depending on which BICEP2/Keck frequency is used, and their shapes vary slightly.

\begin{figure}
  \centering
  \includegraphics[width=\linewidth]{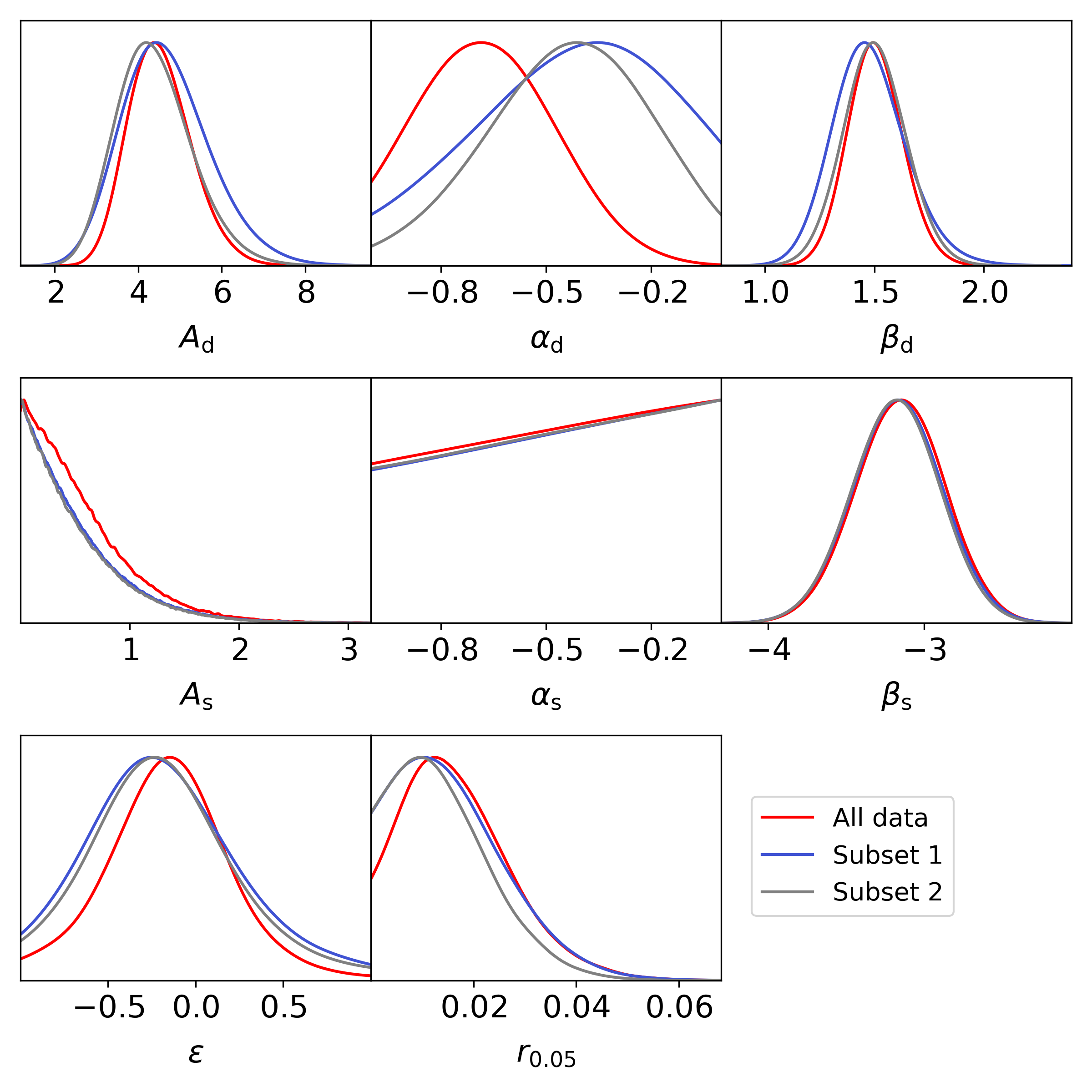}
  \vskip -0.2in
  \caption{Constraints on the tensor-to-scalar ratio, $r$, and the foreground parameters from maps in the BICEP3 sky region (Subset 1, blue), or from both sky regions but neglecting cross-spectra between regions (Subset 2, gray), compared to the full BK18 likelihood (red). The subsets have slightly broader distributions, but $r$ is not strongly affected.}
   \label{fig:r_fg_windows}
\end{figure}

\section{Likelihood}
\label{appendixB}
Following \cite{HamimecheLewis2008,Barkats2014}, the transformed bandpowers, given the data bandpowers $\hat {\mathcal{{D}}}_b$ and model bandpowers ${\mathcal{D}}_b$, are given by
\begin{equation}
    X_b = (\mathcal{D}_b^f)^{1/2}U_b  g(D_b) U_b^\dag(\mathcal{D}_b^f)^{1/2},
\end{equation}
where $\mathcal{D}_b^f$ are fiducial bandpowers calculated from simulations of the signal and noise using fiducial $\Lambda$CDM parameter values. $U_b$ is a matrix of the eigenvectors of $\mathcal{D}_b^{-1/2} \hat{\mathcal{D}}_b \mathcal{D}_b^{-1/2}$, and $D_b$ is a diagonal matrix of the eigenvalues of $\mathcal{D}_b^{-1/2} \hat{\mathcal{D}}_b \mathcal{D}_b^{-1/2}$. The function $g$, given by
\begin{equation}
    g(x) = \rm{sign}(x-1)\sqrt{2(x-\ln x -1)},
\end{equation}
is applied to each element of the diagonal matrix $D_b$ to give $g(D_b)$. 

BK18 uses a suite of signal and noise simulations to calculate the fiducial bandpower covariance matrix $M_{bb'}$, and terms are added to account for the gain and beam width calibration uncertainties.\\

\begin{figure}[t]
  \centering
\vspace{0.2in}

\hspace{-0.2in}
  \includegraphics[width=1.05\linewidth]{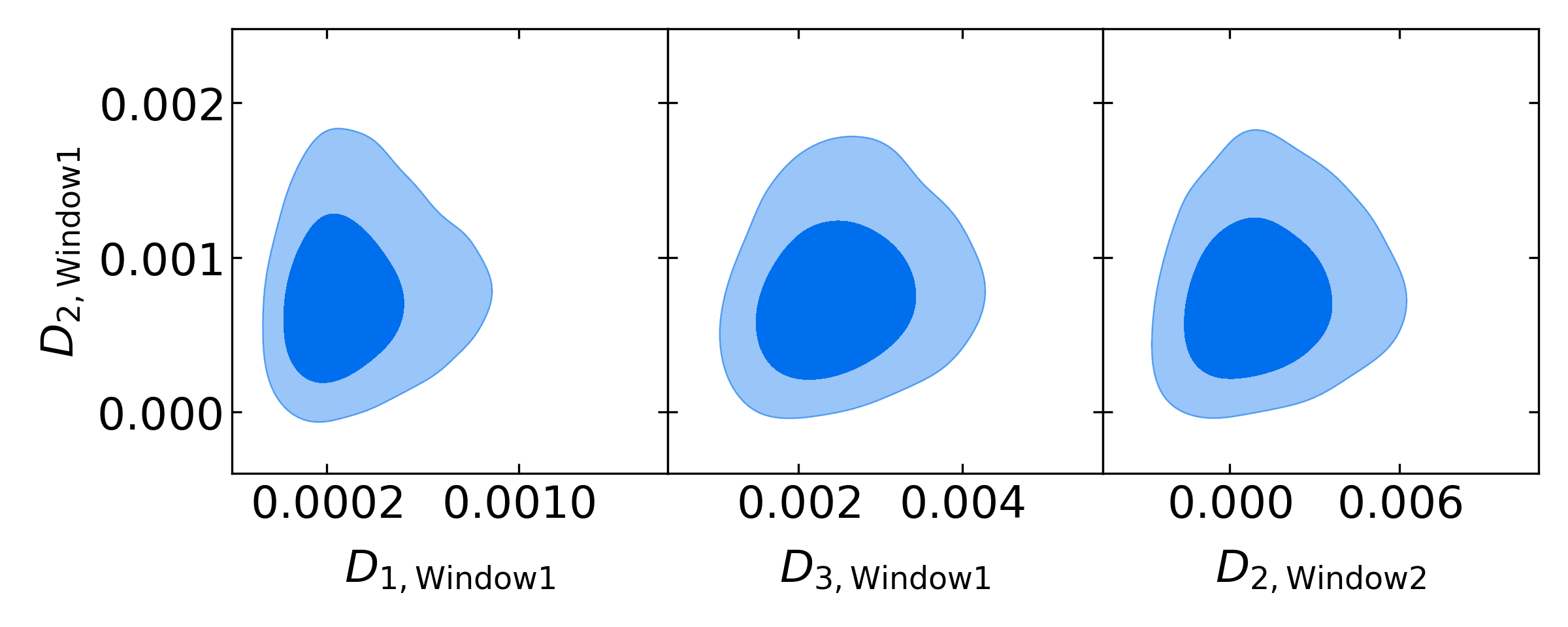}
  \caption{Examples of the $B$-mode bandpowers in the BICEP3 region (Window 1), and smaller BICEP2/Keck region (Window 2), marginalized over foregrounds.}
   \label{fig:Acmb_2D}
\end{figure}

\begin{figure}[b!]
  \centering
  \includegraphics[width=\linewidth]{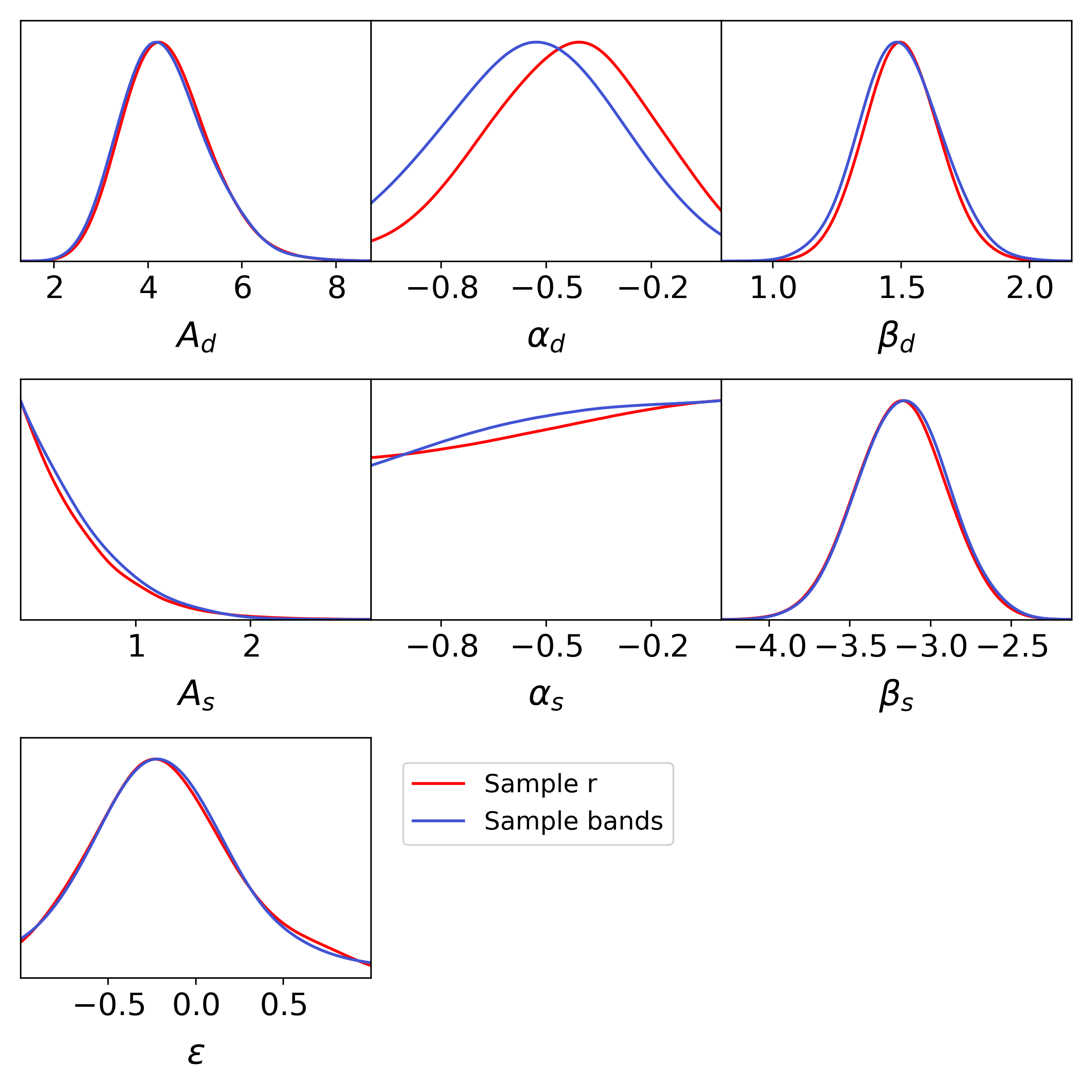}
   \vskip -0.2in
  \caption{Constraints on the seven foreground parameters are consistent whether estimating $r$ or the CMB bandpowers.}
   \label{fig:fg_compare}
\end{figure}

\section{Likelihood tests and estimated bandpowers}
\label{appendixC}

Constraints on the tensor-to-scalar ratio, $r$, and the seven foreground parameters,  are shown in Figure~\ref{fig:r_fg_windows} for the nominal BK18 analysis compared to the two cases we consider: (1) only maps from the BICEP3 region (subset 1), and (2) only spectra computed from pairs of maps in the same footprint, and using BK15 data in the smaller region (subset 2). The constraints are similar in the three cases; there is a $\sim~1\sigma$ shift in the preferred scale dependence of the dust, $\alpha_d$.

Two-dimensional distributions for the second bandpower (with bin center $\ell=72$) and its neighboring bins, estimated from the data in the BICEP3 sky region, are shown in Figure \ref{fig:Acmb_2D} as `Window 1'. The correlation is also shown with the second bandpower estimated from data in the BICEP2/Keck region, marked as `Window 2'. 

The foreground parameters estimated either with $r$ sampled, or the $9\times2$ $B$-mode bandpowers sampled, are shown in Figure \ref{fig:fg_compare}.

\end{document}